\documentclass[preprint,12pt,3p,authoryear]{elsarticle}

\usepackage{type1cm}

\usepackage{color}

\newcommand{\nabo}[2][black]{\textcolor{#1}{#2}}
\newcommand{\new}[2][black]{\textcolor{#1}{#2}}

\newcommand{\neu}[2][black]{\textcolor{#1}{#2}}

\newcommand{\nieuw}[2][black]{\textcolor{#1}{#2}}

\usepackage{dsfont}

\usepackage{amsmath,amssymb}

\usepackage{amsopn}

\usepackage[]{wasysym}

\usepackage{empheq}

\usepackage{blindtext}

\usepackage{graphicx}

\usepackage{booktabs}

\usepackage{lscape}

\usepackage{geometry}

\usepackage{subfigure,caption}

\usepackage{multirow}

\usepackage{bigstrut}
\makeatletter
\def\eqalign#1{\null\,\vcenter{\openup\jot\m@th
  \ialign{\strut\hfil$\displaystyle{##}$&$\displaystyle{{}##}$\hfil
      \crcr#1\crcr}}\,}
\makeatother

\makeatletter
\renewcommand{\paragraph}{%
\@startsection{paragraph}{4}%
{\z@}{1.25ex \@plus 1ex \@minus .2ex}{-1em}%
{\reset@font \normalsize \itshape}%
}
\makeatother

\newcommand{\Secref}[1]{Section~\ref{#1}}
\newcommand{\Figref}[1]{Figure~\ref{#1}}

\newcommand{\Tabref}[1]{Table~\ref{#1}}

\DeclareMathAlphabet{\vektor}{OML}{cmm}{b}{it}

\makeatletter
\DeclareMathSizes{\@xiipt}{\@xiipt}{7.5}{7}
\makeatother

\DeclareSymbolFont{txlargesymbols}{OMX}{txex}{m}{n}
\DeclareSymbolFont{eulargesymbols}{U}{zeuex}{b}{n}
\DeclareMathSymbol{\intop}{\mathop}{txlargesymbols}{"52}

\newcommand{\adaptparm}[7]{#1 & #2 & #3 & #4 & #5 & #6 & #7}
\newcommand{\result}[6]{& #1 & #2 & #3 & #4 & #5 & #6}

\begin{document}

\begin{frontmatter}



\title{Efficient modelling of particle collisions using a non-linear viscoelastic contact force}

\author[tud]{Shouryya~Ray\corref{cor1}}

\cortext[cor1]{Corresponding author}

\ead{Shouryya.Ray@tu-dresden.de}

\author[tud]{Tobias Kempe}

\author[tud]{Jochen~Fr\"{o}hlich}

\address[tud]{Institut f\"{u}r Str\"{o}mungsmechanik, Technische
Universit\"{a}t Dresden, \\ George-B\"{a}hr-Stra{\ss}e 3c, D-01062 Dresden, Germany}

\begin{abstract}
\noindent
In this paper the normal collision of spherical particles is investigated. The particle interaction is modelled in a macroscopic way using the Hertzian contact force with additional linear damping. 
The goal of the work is to develop an efficient approximate solution of sufficient accuracy for this problem which can be used in soft-sphere collision models for Discrete Element Methods and for particle transport in viscous fluids.   
%
%
First, by the choice of appropriate units, the number of governing parameters of the collision process is reduced to one, \new{which is a simple combination of known material parameters as well as initial conditions. It provides} a dimensionless parameter that characterizes all such collisions up to dynamic similitude. 
Next, a rigorous calculation of the collision time and restitution coefficient from the governing equations, in the form of a series expansion in \nieuw{this} parameter \nieuw{is provided}. 
Such a calculation based on first principles is particularly interesting from a theoretical perspective. Since the governing equations present some technical difficulties, the methods employed are also of interest from the point of view of \nieuw{the} analytical technique.
Using further approximations, compact expressions for the restitution coefficient and the collision time are then provided. 
These are used to implement an approximate algebraic rule for computing the desired stiffness and damping in the framework of the adaptive collision model (Kempe \& Fr\"{o}hlich, Journal of Fluid Mechanics, 709: 445-489, 2012).
Numerical tests with binary as well as multiple particle collisions are reported to illustrate the accuracy of the proposed method and its superiority in terms of numerical efficiency.

\end{abstract}


\end{frontmatter}

%
%
%
\section{Introduction and motivation}
\label{sec:Intro}

\noindent
Particle laden flows and their numerical simulation are of \nieuw{major} interest in a wide range of engineering applications as well as in fundamental research. 
A frequently used approach for the simulation of dynamic granular materials is the Discrete Element Method (DEM), cf. \cite{Cundall:1979,Hoomans:1996,poeschel:2005}.
The linear and angular momentum balance of the particles is solved to obtain \nieuw{their} translational and rotational velocity.
The hydrodynamic interaction between particles is often neglected or fluid forces are accounted for by simple empirical correlations and
the particle interaction modelled using macroscopic collision models \nieuw{of various types}. The accurate numerical modelling of the collision process\nieuw{, however,} is crucial for the quality of the simulation in a vast regime of parameters.

Several numerical models for the collision process between particles and for the collision of particles with walls have been developed in the framework of the DEM \citep{KruggelEmden2007,kruggel2008selection,kruggel2009applicable}.
These models can be divided into two groups: hard-sphere models and soft-sphere models. The hard sphere approach \new{does not aim to resolve the details of the collision process. Instead, large time steps are considered and the collisions are treated as} quasi-instantaneous. The post-collisional velocities are calculated from momentum conservation between the states before and after surface contact. The reader is referred to \cite{crowe:2006} for details. Soft-sphere models\nieuw{, on the other hand,} usually require an excessively small time step if physically realistic material parameters are matched.
In the soft-sphere approach the motion of the particles is calculated by numerically integrating the equations of motion of the particles \nieuw{during the collision process} accounting for the contact forces acting on them.
Typical for all soft-sphere models is that very small time steps must be used to ensure that for reasons of stability and accuracy the step size in time is  substantially smaller than the duration of contact. 
The soft-sphere contact forces are usually based on linear and non-linear spring damper models, reviews of which may be found in \cite{KruggelEmden2007,kruggel2008selection,kruggel2009applicable}\nieuw{.}
A commonly used model for time-resolved particle interactions in numerical simulations is a Hertzian contact force in combination with a linear damping \citep{lee1993angle,KruggelEmden2007}.
However, in contrast to linear spring models, no closed solution exists for this equation. One is hence forced to integrate numerically with a very small time step. This issue even leads some researchers to prefer linear spring models which are much easier to evaluate \citep{kruggel2009applicable}.

The discussion whether the linear or the non-linear approach is to be preferred seems unsettled in the community so far \citep{kruggel2009applicable}, and the present paper does not aim to compare these or to advocate one or the other. Instead, the mathematical properties of the equation of damped Hertzian contact are discussed and an efficient engineering approximation is proposed so as to reduce the cost of this model. This can enhance the efficiency of DEMs employing the physically more realistic non-linear approach.

Due to the time-step reduction required by many soft-sphere models, DEM practitioners \nieuw{often} modify the material parameters of the collision model to \nieuw{alleviate} this problem \citep{Hoomans:1996} and allow the use of larger time steps. The idea is to make the collision process softer\nieuw{,} hence longer in time, at the price of increased numerical overlap of particles during collisions. \new{Obviously, the collision process cannot be lengthened arbitrarily. In particular, if the collision time becomes large compared to the free propagation kinematic time scale, the dissipation of kinetic energy may become anomalously small \citep{lit_Luding1994}. Thus, for certain dense multiphase flows, the permissible lengthening of the collision time may only be small. The choice of the stretching factor depends on the regime considered as well as the targeted accuracy and is within the responsibility of the practitioner. Especially for large simulations, even a small increase in step size can lead to significant savings in terms of computational resources and time. The results of the present work would be of use in the implementation of such schemes.}

Except for the linear models, the adjustment of parameters in the soft-sphere model is usually performed by trail and error. This is time consuming and prone to a sub-optimal choice of values.
A current trend in the modelling of particulate flows is that DEMs are enhanced by representations of the viscous effects of the continuous phase around the particles, cf. \cite{Uhlmann:2005,Kempe_2012:JCP}.
In this framework, hard-sphere models are inapplicable as they cannot properly account for the coupling to the surrounding viscous fluid, thus introducing substantial numerical errors \citep{Kempe_2012:JFM}. Here, soft-sphere models are required, with the drawbacks discussed above. As a remedy, a systematic strategy was recently proposed to determine the parameters for a softened contact model \citep{Kempe_2012:JFM}. It is based on imposing the duration of the contact between the particles during collisions according to some external constraint, such as a pre-selected time step. The coefficients in the model are then determined so as to maintain the exact restitution coefficient. This guarantees maximal physical realism under given constraints imposed by computational resources.
The approach, termed Adaptive Collision Time Model (ACTM), was implemented and tested with particles in viscous fluids for single collisions \citep{Kempe_2012:JFM} as well as for multiple simultaneous collisions \citep{kempe:2014_IJMPF}.

Beyond that, the approach is very interesting for pure DEM without viscous fluid, as it provides an automated systematic approach to regularizing the collision process. Another substantial advantage of this approach is that the original physical values of the coefficients can be introduced as bounds so that the original model is obtained again in a regular limit when the collision time is sufficiently reduced. This provides optimal commodity for the user.

\new{It should be noted, that the full collision modelling procedure for particles in viscous flow proposed by \cite{Kempe_2012:JCP}, termed Adaptive Collision Model (ACM), consists not only of the ACTM but also a lubrication model and an Adaptive Tangential Force Model (ATFM). The latter accounts for the effects of tangential forces during surface contact. This issue is not addressed in the present paper, which focuses on the treatment of normal collisions in the ACTM paradigm only.}

In a simulation with many particles, each collision takes place with different velocities of the collision partners. Hence, when imposing \nieuw{the} duration of contact an\nieuw{d the} restitution coefficient, one is forced to select the model coefficients for stiffness and damping for each collision individually.
If no closed solution is available, this requires an iterative procedure as indeed used so far \citep{Kempe_2012:JFM}. In the present paper, this is now improved by providing a direct solution to this problem, based on a systematically controlled approximation. The increased efficiency is demonstrated by suitable test cases and comparison to the original method.

The paper is structured as follows. First, an exact formal solution of the equation of motion for a normal linarly damped Hertzian collision is derived using nonlinear transformations and a parametric series expansion.
Then, a rigorous calculation of the collision time and restitution coefficient is carried out. \nieuw{As a next step}, compact analytical approximations are developed\nieuw{. These} formul\ae{} allow the direct computation of the physically relevant parameters\nieuw{, i.e.} collision time and restitution coefficient\nieuw{,} from the intrinsic material parameters. Afterwards, the inverse problem is addressed. The artificial lengthening of the collision time\nieuw{,} while preserving the restitution coefficient\nieuw{,} requires the computation of the appropriate stiffness and damping. Finally, numerical tests demonstrate the accuracy and efficiency of the proposed algorithm, including test runs in typical engineering settings.


\section{Basic equation of the collision process}
\begin{figure}
\centering
\includegraphics[width=.85\textwidth]{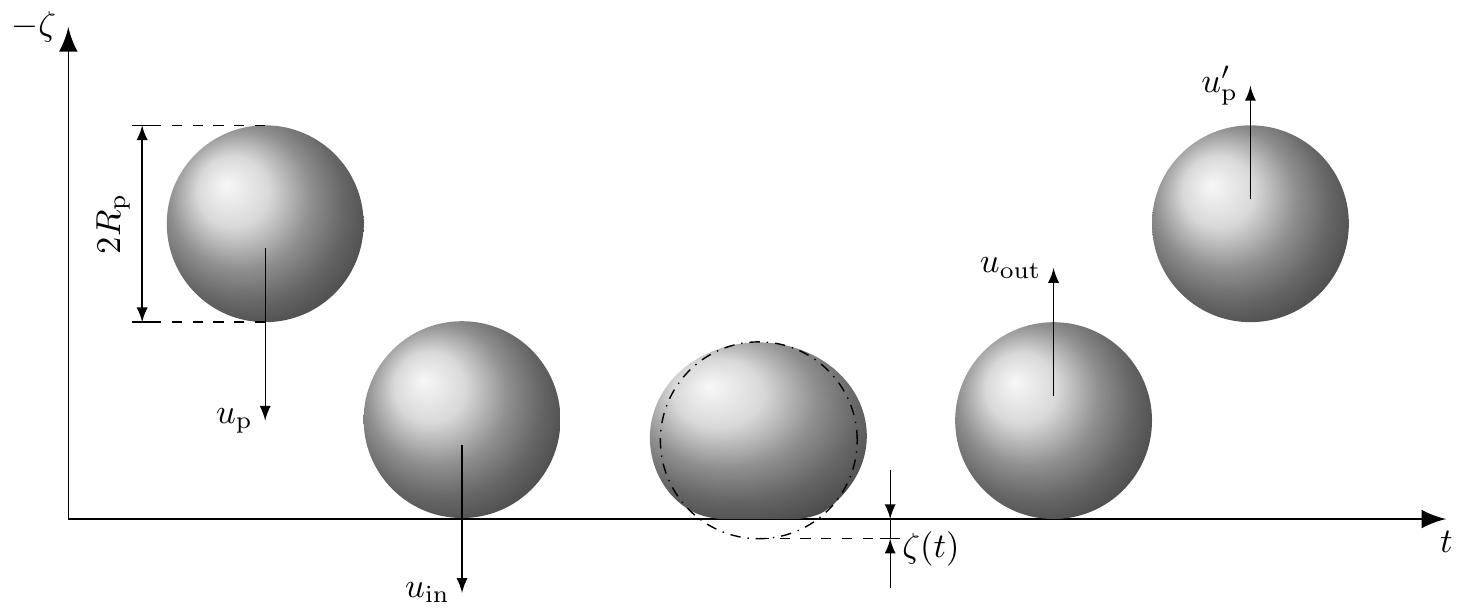}
\caption{Sketch of collision problem with particle position over time}
\label{fig:sketch_collision}
\end{figure}

\noindent 
\nieuw{The situation of a normal particle-wall collision is illustrated in Figure \ref{fig:sketch_collision}}. Particle deformation during contact is represented here by the overlap of the undeformed particle with the collision partner. The equation of motion governing the surface penetration \(\zeta = \zeta(t)\) during the contact phase \nieuw{considered here is \citep{Kempe_2012:JFM}}
\begin{equation} 
m_{\rm p}\ddot{\zeta} = - d\dot{\zeta} - k\zeta^{3/2}
\label{eq:eom}
\end{equation}
with
\(m_{\rm p}\) the mass of the particle\nieuw{, $d$ the damping coefficient} and \(k\) \nieuw{a stiffness parameter, the last two being material properties}. The overdot represents differentiation with respect to time \(t\). The second term \new{on the right-hand side} is the nonlinear restoring force originally derived by \cite{Hertz1882}. The first term on the right hand side of \eqref{eq:eom} corresponds to the damping, which is assumed to be \nieuw{linear}. \nieuw{For $d = 0$ the behaviour is termed ideally elastic.} The initial conditions at the beginning of the collision are \(\zeta(t=0) = 0, \dot{\zeta}(t=0) = u_{\rm in}\).

Equation \eqref{eq:eom} is more conveniently expressed in dimensionless form by defining new variables \(\tau, z\) with \(t = \tau\,t_{*} \text{ and } \zeta = z\,u_{\rm in} t_{*}\), which is tantamount to fixing the characteristic unit of velocity for the system as \(u_{\rm in}\) and choosing \nieuw{$t_*$ as} the (at present arbitrary) unit of time.
The first and second derivatives of $\zeta$ can \neu{then} be expressed as
\begin{equation}
	\dot{\zeta}(t)=u_{in}\:\neu{\dot{z}(\tau)} \hspace{0.5cm}
\end{equation}
and
\begin{equation}
	\ddot{\zeta}(t)=\frac{u_{in}}{t_*}\:\neu{\ddot{z}(\tau)} \hspace{0.5cm}.
\end{equation}
\neu{Here and in all further cases of occurrence, the overdot denotes, when applied to a dimensionless quantity, differentiation with respect to the dimensionless time~\(\tau\)\nieuw{, unless stated otherwise}. Dividing by the characteristic unit of force \(m_{\rm p} u_{\rm in}/t_{*}\), \nieuw{one} obtain\nieuw{s} the dimensionless equation
\begin{equation}
\ddot{z} + \frac{t_{*}d}{u_{\rm in} m_{\rm p}}\dot{z} + \frac{k u_{\rm in}^{1/2}}{m_{\rm p}t_{*}^{5/2}}z^{3/2} = 0.
\end{equation}
Choosing the unit of time \(t_{*}\) as
\begin{align}
t_{*} & = \sqrt[5]{\frac{m_{\rm p}^2}{k^2 u_{\rm in}}} \label{eq:scale-t} \hspace{0.5cm},
\end{align}
\nieuw{yields} the following equation for \(z(\tau)\):
\begin{equation}
\ddot{z} + 2 \lambda \dot{z} + z^{3/2} = 0 \hspace{0.5cm},
\label{eq:eom_nondim}
\end{equation}
with initial conditions \(z(\tau=0) = 0, \dot{z}(\tau=0) = 1\) and the parameter
\begin{align}
\lambda & = \frac{1}{2} \frac{t_* d}{m_p}=\frac{1}{2}\frac{d}{m_{\rm p}}{\sqrt[5]{\frac{m_{\rm p}^2}{k^2u_{\rm in}}}}\hspace{0.25cm}.
\label{eq:scale-lambda}
\end{align}
}

Following \cite{Kempe_2012:JFM}, the collision time \(T_{\rm
c}\)\nieuw{, i.e.} \(\tau_{\rm c}\) \nieuw{in the non-dimensional setting,} is obtained as the strictly positive root of \(\zeta\), i.e.
the point of time when the surface penetration of the particle-wall system
returns to zero. The restitution coefficient is then conveniently defined as
\begin{equation}
e_{\rm dry} = -\frac{u_{\rm out}}{u_{\rm in}} = -\frac{\dot{\zeta}(t=T_{\rm c})}{u_{\rm in}} =
-\dot{z}(\tau=\tau_{\rm c})\nieuw{\;.}
\end{equation}
The latter expression makes clear that the only way the restitution
coefficient may depend on material and other input parameters is as a function
of the parameter \(\lambda\) introduced above. Physically, one may
interpret this as follows: All linearly damped normal Hertzian
particle-wall collisions are, up to a characteristic parameter,
similar. This characteristic
parameter can be the restitution coefficient \(e_{\rm dry}\), which is
easy to determine experimentally but cannot be calculated trivially from known
material properties and initial conditions. The other option, introduced
here, is \new{to choose \(\lambda\) as the characteristic parameter}, which cannot be measured directly but is readily
calculated. At the heart of the present work is the convenient
analytical conversion between the two. The case \(\lambda=0\) corresponds to the undamped case considered by \cite{Hertz1882}, which is simply a conservative system, i.e. \(e_{\rm dry} = 1\), and is integrated readily. Furthermore, \cite{Hertz1882} gave the collision time as
\begin{equation}
T_{\rm c,0} = \frac{2\sqrt{\pi}\,\Gamma\!\left(\frac{7}{5}\right)}{\Gamma\!\left(\frac{9}{10}\right)}\sqrt[5]{\frac{25m_{\rm_p}^2}{16k^2u_{\rm in}}}\:,
\end{equation}
with \(\Gamma\) denoting the Gamma function. \new{Evaluation of all factors to four significant figures} yields \(\tau_{\rm c,0} = 3.218\) in dimensionless form.


\section{Exact formal solution of the equation of motion}
\label{sec:exsol}

\subsection{Introduction}
\noindent
The purpose of this section is to calculate rigorously from first principles, i.e. starting from the equation of motion \eqref{eq:eom_nondim} and using mathematically justifiable techniques, the physical parameters pertaining to the characterization of collisions in the chosen model. Such results are not only useful for the purpose of theoretical investigation, but are also instructive from the perspective of the analytical technique. Since \nieuw{these results} are, to the best of our knowledge, not available in the existing literature, \nieuw{their derivation is presented} in the following.

Due to the fractional power of \(z\) in \eqref{eq:eom_nondim} and the initial condition \(z(\tau=0) = 0\), a naive power series solution for \(z\) in \(\tau\) is not possible. Furthermore, it is not desirable at all to have a series representation in \(t\), since it is not a variable that is naturally small. On the other hand, especially in practical applications, very low values of \(e_{\rm dry}\) are unlikely to be of interest. By consequence, \(\lambda\) is a parameter that may be assumed small for practical purposes. Indeed, numerical examples show that if a lower bound of \(0.4\) \nieuw{is imposed} on \(e_{\rm dry}\), \(\lambda \leqslant 0.2\) may be safely assumed. \nieuw{In addition, as remarked earlier, the case $\lambda = 0$ has already been studied in detail and is well-understood \citep{Hertz1882}.} Thus, in the following, the solution of the equation of motion is \nieuw{most conveniently} expressed using series \nieuw{expansions} in \(\lambda\).

\subsection{Solving the equation of motion}
\noindent
It will be found that the relevant mathematical operations are \nieuw{substantially} facilitated when the problem is transformed to phase space \((z,\dot{z})\), \new{as visualised in \Figref{fig:zt_ph}. With this approach,} the dimensionless velocity can be interpreted as a function of the penetration, i.e. \(\dot{z} = \dot{z}(z)\). Using the relation 
$\ddot{z}=\left.{\rm d}\dot{z} \middle/ {\rm d}\tau\right. = \left.{\rm d}z \middle/ {\rm d}\tau\right. \left.{\rm d}\dot{z} \middle/ {\rm d}z\right. = \dot{z} \left.{\rm d}\dot{z} \middle/ {\rm d}z\right. $
the equation of motion expressed in these new variables reads 
\begin{equation}
\dot{z}\frac{\partial \dot{z}}{\partial z} + 2\lambda\dot{z} + z^{3/2} = 0.
\label{eq:phasespace}
\end{equation}
In passing, observe that if the damping term were absent, this would reduce to an exact differential, with the solution given by the level curve of a potential function which \nieuw{is} readily interpret\nieuw{ed} as the conserved total energy of the system. \nieuw{In fact, the} energy method was \nieuw{the procedure} \nieuw{followed} by \cite{Hertz1882} when \nieuw{deriving the solution for} the undamped case. The analysis here, however, needs to be more involved. \nieuw{It is now convenient to define}
\begin{equation}
\dot{z} =: \begin{cases}
\sqrt{v_{+}} & \text{in} \\
-\sqrt{v_{-}} & \text{out}
\end{cases}
\label{def_v}
\end{equation}
where `in' refers to the part of the collision during which the particle is compressed and the motion is directed into the wall, while `out' describes the outward motion of the particle (Figure \ref{fig:zt_ph}). \neu{By the chain rule, \nieuw{one has} \(\dot{z}{\rm d}\dot{z} = \frac{1}{2}{\rm d}v_{\pm}\), where the choice of \(+\) and \(-\) is determined by \eqref{def_v}.}
Finally, \nieuw{it is helpful to define} \(y := \sqrt{z}\), whence \(\neu{{\rm d}z = 2y\,{\rm d}y}\). Inserting these new variables into \eqref{eq:phasespace} \nieuw{yields} the initial value problems \nieuw{of the two phases of motion:}
\begin{align}
\frac{\partial v_{+}}{\partial y} + 8\lambda y \sqrt{v_{+}} + 4y^4 & = 0 \qquad \qquad v_{+}(y = 0) = 1 \label{eq:govv+} \\
\frac{\partial v_{-}}{\partial y} - 8\lambda y \sqrt{v_{-}} + 4y^4 & = 0 \qquad \qquad v_{-}(y = 0) = e_{\rm dry}^2\hspace{0.2cm}. \label{eq:govv-}
\end{align}
For \(\lambda\) assumed small, \nieuw{the unknowns are now expressed in form of a series in powers of $\lambda$}:
\begin{align}
v_{+} = \sum_{m=0}^{\infty}v_{+,m}\lambda^{m} \qquad \qquad  v_{+,m} = \frac{1}{m!}\!\left.\frac{\partial^m v_+}{\partial \lambda^m}\right|_{\lambda=0} \\
v_{-} = \sum_{m=0}^{\infty}v_{-,m}\lambda^{m} \qquad \qquad v_{-,m} = \frac{1}{m!}\!\left.\frac{\partial^m v_-}{\partial \lambda^m}\right|_{\lambda=0}
\end{align}
Differentiating \(m\) times with respect to \(\lambda\) and evaluating at \(\lambda=0\) successively yields the equations determining \(v_{\pm,m}\) for \(m \geqslant 0\). For \(m=0\), this gives
\begin{equation}
v_{\pm,0} = 1 - \tfrac{4}{5}y^5
\label{eq:zerothorder}
\end{equation}
\neu{\nieuw{and for} \(m\geqslant 1\), 
\begin{equation}
m!\,\frac{\partial v_{\pm,m}}{\partial y} = \mp 8y\left.\frac{\partial^m}{\partial \lambda^m}\lambda \sqrt{v_{\pm}}\right|_{\lambda=0}
\label{eq:vmv1}
\end{equation}
The right-hand side of \eqref{eq:vmv1} may be further simplified using \nieuw{the} Leibniz formula,
\begin{equation}
\left.\frac{\partial^m}{\partial \lambda^m}\lambda\sqrt{v_{\pm}}\right|_{\lambda=0} = \sum_{i=0}^{m}\binom{m}{i}\left.\frac{\partial^i\lambda}{\partial \lambda^i}\right|_{\lambda=0}\left.\frac{\partial^{m-i}\sqrt{v_{\pm}}}{\partial \lambda^{m-i}}\right|_{\lambda=0}\;.
\end{equation}
Only the term \(i=1\) survives, so that} the higher order corrections are given by
\begin{align}
v_{+,m} &= -\frac{8}{(m-1)!}\int_{0}^{y}\!\left.\frac{\partial^{m-1} \!\sqrt{v_{+}}}{\partial \lambda^{m-1}}\right|_{\lambda=0}y^\prime {\rm d}y^\prime\;;\label{eq:v+m-master} \\
v_{-,m} &= \frac{1}{m!}\!\left.\frac{\partial^m e_{\rm dry}^2}{\partial \lambda^m}\right|_{\lambda=0} + \frac{8}{(m-1)!}\int_{0}^{y}\!\left.\frac{\partial^{m-1} \!\sqrt{v_{-}}}{\partial \lambda^{m-1}}\right|_{\lambda=0} y^\prime{\rm d}y^\prime \;.
\label{eq:v-master}
\end{align}
Since the \((m-1)\)-st derivative of \(v_{\pm}\) at \(\lambda=0\) can only involve \(v_{\pm,0},\ldots,v_{\pm,m-1}\), equations \eqref{eq:v+m-master} and \eqref{eq:v-master} yield well-defined recursions for \(v_{\pm,m}\). Furthermore, it is readily seen by mathematical induction, that \(v_{\pm,m}\) have a convergent Taylor series expansion around \(y=0\), and that the convergence radius is \((5/4)^{1/5}\). From conservation of energy\nieuw{,} however, it can be deduced that \(z \leqslant (5/4)^{2/5}\), i.e. \(v_{\pm,m}\) can be represented by a power series on the whole domain of interest. 
A physical interpretation of these terms is discussed in \ref{app:PhysInt}. Note also that derivatives of \(e_{\rm dry}\) appear explicitly in the expressions for \(v_{-,m}\). This does not imply that \(e_{\rm dry}\) needs to be known \emph{a priori}. Rather, it will be seen later that \(e_{\rm dry}\) is uniquely determined by certain conditions that need to be fulfilled in order for the trajectory to be continuous in time and space.

To find a connection to the time domain, one may exploit the fact that \(\dot{z}\,{\rm d}\tau = {\rm d}z\) to find
\begin{equation}
\tau = \tau(y) = \begin{cases}\tau_{+}(y) := \displaystyle\int_{0}^y \frac{2\displaystyle{\rm d}y^\prime y^\prime}{\displaystyle\sqrt{v_{+}}} & \qquad \text{in}\:, \\
\tau_{-}(y) := \tau_{+}(\hat{y}) + \displaystyle\int_0^y\frac{2\displaystyle{\rm d}y^\prime y^\prime}{\displaystyle\sqrt{v_{-}}} & \qquad \text{out}\:,
\end{cases}
\label{eq:explsol}
\end{equation}
with \(\hat{y}^2 = \hat{z}\) the maximum surface penetration. This solves the problem in the classical sense. In order to obtain an explicit relation of the form \(z = z(\tau)\), one would have to take the functional inverse and square the resultant expression. However, since it does not contribute to the calculation of the relevant physical properties \(\hat{z}, e_{\rm dry}\) and \(\tau_{\rm c}\), that part of the problem, though utterly non-trivial, will not be addressed \nieuw{here}.

\subsection{Calculation of the physical parameters} 
\label{sec:exsol:physparms}
\noindent
With the above solution \eqref{eq:v+m-master}--\eqref{eq:explsol} of the equation of motion at hand, it is now possible to calculate the significant physical quantities. Although the two primary parameters of interest are the restitution coefficient \(e_{\rm dry}\) and collision time \(\tau_{\rm c}\), the calculation of the maximum penetration \(\hat{z}\) is required as an intermediate step \nieuw{and is therefore addressed first}.
From the definition of $\dot{z}$, it follows that \(\dot{z} \overset{!}{=} 0\), \nieuw{which is equivalent to} the more tractable condition
\begin{equation}
v_{+}(y=\hat{y}) = 0\:,
\label{eq_necesmaxpen}
\end{equation}
with \(\hat{y}^2 = \hat{z}\) the maximum penetration. This equation uniquely determines \(\hat{z}\) \nieuw{in terms of} the smallest positive real root of $v_+$. Since the solution trajectory is required to be continuous, the velocity on both the inward and outward trajectory must go to zero at the same value \nieuw{of} \(y\)\nieuw{,} \nieuw{y}ielding the condition
\begin{equation}
v_{-}(y=\hat{y}) = 0 \: .
\end{equation}
This uniquely determines \(e_{\rm dry}\). Finally, the collision time can be computed directly as
\begin{equation}
\tau_{\rm c} = \tau_{-}(\hat{y})
\end{equation}

Obviously, each of the above quantities depend on \(\lambda\). For \(\lambda=0\), the classical case of the undamped Hertzian restoring force is obtained, for which the solutions are \citep{Hertz1882}
\begin{align}
\hat{y}_{0} := \hat{y}(\lambda=0) & = \left(\tfrac54\right)^{\!1/5}\:,  \\
e_{\rm dry}(\lambda=0) & = 1\:, \\
\tau_{\rm c,0} := \tau_{\rm c}(\lambda=0) & = 3.218\:.
\end{align}
For \(\lambda\) small, therefore, it is natural to develop the solutions in terms of a power series in \(\lambda\) around $\lambda = 0$, whereby the respective governing equations are successively differentiated at \(\lambda=0\) using implicit differentiation, with the chain rule used to evaluate the Taylor coefficients.
\new{\subsection{Explicit computation of first-order corrections}
\label{subsec_expliccomput1order}}
\noindent \new{Following the procedure specified above, it should, in principle, be possible to calculate the correction terms to arbitrarily high order, which naturally becomes more and more involved with increasing order. Here, the explicit computation of the first-order corrections is presented for illustration.
}
\paragraph{Maximum surface penetration}
To zeroth order, as established above, condition \eqref{eq_necesmaxpen} yields the result \(\hat{y} = \sqrt[5]{5/4} + \mathcal{O}(\lambda)\). This is the same solution as one would estimate using conservation of energy, since to zeroth order in \(\lambda\), the dissipation of energy, i.e. work done by the damping force is negligible. Differentiating \(v_{+}(\hat{y}) = 0\) once with respect to \(\lambda\) and using the chain rule to account for the fact that \(\hat{y}=\hat{y}(\lambda)\), one arrives upon evaluation at $\lambda = 0$ the condition 
\[
\left.\frac{\partial v_{+,0}}{\partial y}\right|_{y = \hat{y}_0}\left.\frac{\partial \hat{y}}{\partial \lambda} \right|_{\lambda=0} + v_{+,1}(y=\hat{y}_0) = 0\;,
\] \nieuw{or}
\begin{equation}
-4\hat{y}_0^4\left. \frac{\partial \hat{y}}{\partial \lambda}\right|_{\lambda=0} - 8\int_{0}^{\hat{y}_0}\! \sqrt{1 - \tfrac{4}{5}y^5}\,y\,{\rm d}y = 0 
\end{equation}
Carrying out the integration yields
\begin{equation}
\hat{y} = \sqrt[5]{\frac{5}{4}} -\frac{1}{2}\sqrt[5]{\frac{16}{25}}\frac{\sqrt{\pi}\,\Gamma\!\left(\frac{7}{5}\right)}{\frac{9}{10}\Gamma\!\left(\frac{9}{10}\right)}\lambda + \mathcal{O}(\lambda^2)\;.
\end{equation}

\paragraph{Restitution coefficient} Recall that imposing continuity on the solution yields the condition \(v_{-}(\hat{y}) = v_{+}(\hat{y}) = 0\). Again, to zeroth order one finds \(\left.e_{\rm dry}\right|_{\lambda=0} = 1\), which is an obvious consequence of conservation of energy (the dissipation through damping being first order and higher in \(\lambda\)). Differentiating the expression for \(v_{-}(\hat{y})\) \nieuw{yields} 
\[
\left.\frac{\partial v_{-,0}}{\partial y}\right|_{y=\hat{y}_0} \left.\frac{\partial \hat{y}_0}{\partial \lambda}\right|_{\lambda=0} + v_{-,1}(y=\hat{y}_0)=0
\] and hence
\begin{equation}
\left.\frac{\partial e_{\rm dry}}{\partial \lambda}\right|_{\lambda=0} = 2\hat{y}_0^4 \left.\frac{\partial \hat{y}}{\partial \lambda}\right|_{\lambda=0} - 4\int_{0}^{\hat{y}_0}\!\sqrt{1 - \tfrac{4}{5}y^5}\,y\,{\rm d}y = -8\int_{0}^{\hat{y}_0}\!\sqrt{1 - \tfrac{4}{5}y^5}\,y\,{\rm d}y \;.
\end{equation}
Carrying out the integration yields
\begin{equation}
e_{\rm dry} = 1 - 2\sqrt[5]{\frac{25}{16}}\frac{\sqrt{\pi}\,\Gamma\!\left(\frac{7}{5}\right)}{\frac{9}{10}\Gamma\!\left(\frac{9}{10}\right)}\lambda + \mathcal{O}(\lambda^2)
\label{eq:beauty_e-dry_deriv}
\end{equation}

\noindent\paragraph{Collision time} The collision time requires a computation of the integral
\begin{equation}
\tau_{\rm c} = \int_{0}^{\hat{y}}\!\left(\frac{1}{\sqrt{v_{+}}} + \frac{1}{\sqrt{v_{-}}}\right) 2y{\rm d}y
\end{equation}
Since \(v_{\pm}=v_{\pm,0} + v_{\pm,1}\lambda + \cdots\), \nieuw{this gives}
\begin{align}
\tau_{\rm c} &= \int_{0}^{\hat{y}}\!\left[\frac{1}{\sqrt{v_{+,0}}}\left(1 + \dfrac{v_{+,1}}{v_{+,0}}\lambda + \mathcal{O}(\lambda^2)\right)^{-1/2} + \frac{1}{\sqrt{v_{-,0}}}\left(1 + \dfrac{v_{-,1}}{v_{-,0}}\lambda + \mathcal{O}(\lambda^2)\right)^{-1/2}\right] 2y\,{\rm d}y\nonumber\\
&= \int_{0}^{\hat{y}}\left[\frac{2}{\sqrt{1 - \frac{4}{5}y^5}} + \sqrt[5]{\frac{25}{16}}\frac{\sqrt{\pi}\,\Gamma\!\left(\frac{7}{5}\right)}{\frac{9}{10}\Gamma\!\left(\frac{9}{10}\right)}\frac{2\lambda}{\left(\sqrt{1 - \frac{4}{5}y^5}\right)^{\!\!3}} + \mathcal{O}(\lambda^2)\right]2y\,{\rm d}y
\end{align}
using \eqref{eq:zerothorder}, \eqref{eq:v+m-master}, \eqref{eq:v-master} and \eqref{eq:beauty_e-dry_deriv}, as well as the Taylor series representation \nieuw{of} \((1+x)^{-1/2}\).
\nieuw{For} \(\lambda=0\), the integral is straightforward, \nieuw{yielding} the same result as obtained by \cite{Hertz1882}
\begin{equation}
\tau_{\rm c} = \frac{2\sqrt{\pi}\,\Gamma\!\left(\frac{7}{5}\right)}{\Gamma\!\left(\frac{9}{10}\right)}\sqrt[5]{\frac{25}{16}} + \mathcal{O}(\lambda)\:.
\end{equation}
For the first-order correction, differentiating with respect to \(\lambda\) using differentiation under the integral sign (Leibniz rule) and setting \(\lambda=0\) gives
\begin{equation}
\left.\frac{\partial \tau_{\rm c}}{\partial \lambda}\right|_{\lambda=0} = -\sqrt[5]{\frac{16}{25}}\frac{\sqrt{\pi}\,\Gamma\!\left(\frac{7}{5}\right)}{\frac{9}{10}\Gamma\!\left(\frac{9}{10}\right)}\left.\frac{2y}{\sqrt{1 - \frac{4}{5}y^5}}\right|_{y=\hat{y}_0} + \sqrt[5]{\frac{25}{16}}\frac{\sqrt{\pi}\,\Gamma\!\left(\frac{7}{5}\right)}{\frac{9}{10}\Gamma\!\left(\frac{9}{10}\right)}\int_{0}^{\hat{y}_0}\!\frac{4y\,{\rm d}y}{\left(\sqrt{1 - \frac{4}{5}y^5}\right)^{\!\!3}}\;.
\label{eq:tau-nearlythere}
\end{equation}
Both terms are divergent. The first term on the right hand side has a simple pole at \(y=\hat{y}_0\) and the integrand in the second term diverges too fast for \(y\rightarrow\hat{y}_0\). The result can, however, be regularized by calculating the integral first for general \(y<\hat{y}_0\) and then taking the limit of the whole expression for \(y\rightarrow\hat{y}_0\), i.e.:
\begin{equation}
\new{
\left.\frac{\partial \tau_{\rm c}}{\partial \lambda}\right|_{\lambda=0} = \lim\limits_{y \uparrow \hat{y}_0}\left[-\sqrt[5]{\frac{16}{25}}\frac{\sqrt{\pi}\,\Gamma\!\left(\frac{7}{5}\right)}{\frac{9}{10}\Gamma\!\left(\frac{9}{10}\right)}\frac{2y}{\sqrt{1 - \frac{4}{5}y^5}} + \sqrt[5]{\frac{25}{16}}\frac{\sqrt{\pi}\,\Gamma\!\left(\frac{7}{5}\right)}{\frac{9}{10}\Gamma\!\left(\frac{9}{10}\right)}\int_{0}^{y}\!\frac{4y^\prime\,{\rm d}y^\prime}{\left(\sqrt{1 - \frac{4}{5}{y^\prime}^5}\right)^{\!\!3}}\right]\;.
}
\end{equation}
\new{Evaluating the integral in the second term of the right-hand side yields a contribution $\sim y^2(1 - \tfrac{4}{5}y^5)^{-1}$ and a second one involving a hypergeometric function. One arrives at the following expression:
\begin{equation}
\left.\frac{\partial \tau_{\rm c}}{\partial \lambda}\right|_{\lambda=0} = \lim\limits_{y \uparrow \hat{y}_0} \frac{\sqrt{\pi}\,\Gamma\!\left(\frac{7}{5}\right)}{\frac{9}{10}\Gamma\!\left(\frac{9}{10}\right)}\left[\frac{1}{\hat{y}_0^2}\frac{2y\left(y/\hat{y}_0 - 1\right)}{\sqrt{1 - \left(y/\hat{y}_0\right)^5}} + \frac{2\hat{y}_0^2}{5} y^2 {}_2 F_1\!\left(\tfrac{2}{5},\tfrac{1}{2};\tfrac{7}{5};\left(y/\hat{y}_0\right)^5\right)\right]\;,
\end{equation}
where ${}_2 F_1(\cdot)$ denotes the (ordinary) hypergeometric function \cite[cf.][Chapter 15, pp. 555]{AbraSteg}. Note that, wherever opportune, the factors have been expressed in terms of $\hat{y}_0$ for ease of manipulation. The first term in the limit can be evaluated using de l'H\^{o}pital's rule and is found to vanish for $y \rightarrow \hat{y}_0$. The second term remains regular in the limit and can be simply evaluated at $y = \hat{y}_0$, yielding
}
\begin{equation}
\left.\frac{\partial \tau_{\rm c}}{\partial \lambda}\right|_{\lambda=0} =  \frac{2}{5}\hat{y}_0^4\frac{\sqrt{\pi}\,\Gamma\!\left(\frac{7}{5}\right)}{\frac{9}{10}\Gamma\!\left(\frac{9}{10}\right)} \;{}_2F_{1}\!\left(\tfrac{2}{5},\tfrac{1}{2};\tfrac{7}{5};1\right)\;.
\end{equation}
This can be simplified using Gamma functions, so that one finally has
\begin{equation}
\tau_{\rm c} = \frac{2\sqrt{\pi}\,\Gamma\!\left(\frac{7}{5}\right)}{\Gamma\!\left(\frac{9}{10}\right)}\sqrt[5]{\frac{25}{16}} + \frac{\sqrt[5]{4\cdot 5^4}}{9}\!\left(\!\frac{\sqrt{\pi}\,\Gamma\!\left(\frac{7}{5}\right)}{\Gamma\!\left(\frac{9}{10}\right)}\!\right)^{\!\!\!2}\lambda + \mathcal{O}(\lambda^2)\;.
\end{equation}
\new{\subsection{Summary of analytical results}}
\noindent \new{To summarize, the upshot of the theoretical calculations presented in \Secref{sec:exsol} are expansions of the relevant physical quantities in terms of the (small) parameter $\lambda$. Thus, for the square root of the maximum penetration}
\begin{empheq}{alignat=4}
\sqrt{\hat{z}} &= \hat{y}_{0} &{}- 0.748 \lambda & {}+ 0.578\lambda^2 + \mathcal{O}(\lambda^3)\;.  \label{eq:beauty_hatz}
\end{empheq}
This finally yields:
\begin{empheq}{alignat=4}
e_{\rm dry} &= 1 &{}- 3.576\lambda & {}- 5.131\lambda^2 + \mathcal{O}(\lambda^3) \label{eq:beauty_e-dry}\\
\tau_{\rm c} &= \tau_{\rm c,0} &{}+ 1.152 \lambda & {}+ \mathcal{O}(\lambda^2) \label{eq:beauty_tau-c}
\end{empheq}
where the special functions and other constant numerical factors were evaluated to four significant figures.
\vspace{1.5em}
\section{Approximate method and results}
\label{sec:approx}
\noindent
While the solution \eqref{eq:beauty_e-dry}, \eqref{eq:beauty_tau-c} derived in the previous section is formally correct, calculating the necessary quantities can be costly as higher-order terms must be included for sufficient accuracy.
If, however, terms higher than quadratic are involved the expressions cannot be conveniently inverted. \nieuw{Indeed, the inversion of \eqref{eq:beauty_e-dry} and \eqref{eq:beauty_tau-c} cannot be carried out by algebraic means if quintic or higher-order terms are present.} The inversion could be achieved using the theta function, but this is a non-elementary special function and costly to evaluate. In the following, a more elegant procedure is developed, based on the fact that the expressions of \(e_{\rm dry}(\lambda)\) or \(\tau_{\rm c}(\lambda)\) can be resolved for \(\lambda\).

To this end, consider the following ansatz with $A,B,C$ denoting free parameters whose values are to be determined appropriately:
\begin{align}
e_{\rm dry} & = \exp\!\left(-\frac{\alpha \lambda\tau_{\rm c,0}}{\sqrt{1 - C\lambda}}\right)\;, \label{eq:edry-approx}\\
\tau_{\rm c} & = \frac{\tau_{\rm c,0}}{\sqrt{1 - A\lambda - B\lambda^2}}\;\,. \label{eq:tau-c-ansatz}
\end{align}
The reason one may expect \eqref{eq:edry-approx}, \eqref{eq:tau-c-ansatz} to be a more convenient choice for describing the required quantities than the simple power series in \eqref{eq:beauty_e-dry} and \eqref{eq:beauty_tau-c} is because they mimic to a great extent the relationships that are valid in the case of the damped harmonic oscillator which is obtained if the power $\frac32$ is replaced by unity in \eqref{eq:eom_nondim}. The equation of motion of the harmonic oscillator is readily solved in closed form and yields the behaviour $\log e_{\rm dry} \propto -\delta \tau_{\rm c}$ and $\tau_{\rm c} \propto (1 - \delta^2)^{-1/2}$, where $\delta$ corresponds to the damping ratio.

The value of most of the free parameters in \eqref{eq:edry-approx} and \eqref{eq:tau-c-ansatz} can be determined directly from the solution \eqref{eq:beauty_e-dry} and \eqref{eq:beauty_tau-c} found in \Secref{sec:exsol}. Differentiating \eqref{eq:edry-approx} at $\lambda = 0$ yields the two equations
\begin{align}
\left.\frac{\partial e_{\rm dry}}{\partial \lambda}\right|_{\lambda = 0} & = \alpha \tau_{\rm c,0}\;, \\
\left.\frac{\partial^2 e_{\rm dry}}{\partial \lambda^2}\right|_{\lambda = 0} & = \alpha \tau_{\rm c,0} \left(\alpha \tau_{\rm c,0} - C\right)\;.
\end{align}
The derivatives appearing on the left-hand side are given by \eqref{eq:beauty_e-dry}, whence one obtains
\begin{equation}
\alpha = 1.111 \quad\text{and}\quad C = 0.744 \;, \label{eq:valueaC}
\end{equation}
when evaluated to four significant figures.
Likewise, differentiating \eqref{eq:tau-c-ansatz} yields
\begin{equation}
\left.\frac{\partial \tau_{\rm c}}{\partial \lambda} \right|_{\lambda=0} = \tfrac{1}{2}\tau_{\rm c,0}A \: ,
\end{equation}
so that
\begin{equation}
A = 0.716\:. \label{eq:valueA}
\end{equation}
The method is difficult to employ for determining the remaining free parameter $B$, because it requires knowledge of the second-order correction to $\tau_{\rm c}$, which in turn necessitates the summation of divergent terms (in the form of derivatives of terms like $(1 - \frac{4}{5}y^5)^{-1/2}$ with respect to $y$ at $y = \hat{y}_0$) and improper divergent integrals. This appears, in fact, to be a characteristic shared by corrections to the collision time of all non-zero orders. For the first-order correction, it was possible to carry out the regularization of the result by hand, because the indefinite integral underlying the divergent improper integral could be formulated in closed form and the cancellation of the divergent terms could be done away with in a straightforward manner. For higher order corrections, the integrals become increasingly involved; so much so, that even for the second-order term, the integration could not be carried out by hand. The calculation of $B$ by self-consistent methods is, hence, a highly non-trivial task which may be addressed in a future investigation. 

For the present study, it shall suffice to obtain an approximation by a method similar to data-fitting. To this end, the dimensionless collision time is determined from numerically calculated solutions of the governing equation \eqref{eq:eom_nondim} for several values of $\lambda$. The numerical integration was carried out using a Runge-Kutte method with an adaptive step size, employing the routine \texttt{ode45} of \texttt{MATLAB}\texttrademark{}. Equation \eqref{eq:tau-c-ansatz} predicts that $\left(\tau_{\rm c,0} \middle/ \tau_{\rm c}\right)^2 + A\lambda$ is a linear function of $\lambda^2$. Linear regression of the same yields
\begin{equation}
B = 0.830\:. \label{eq:valueB}
\end{equation}

The result of this section now is the approximation \eqref{eq:edry-approx}, \eqref{eq:tau-c-ansatz} with the parameter values given in \eqref{eq:valueaC}, \eqref{eq:valueA} and \eqref{eq:valueB}. This shall serve as a convenient alternative to the series expansions \eqref{eq:beauty_e-dry}, \eqref{eq:beauty_tau-c} for the purposes of the subsequent sections.

In spite of the simple nature of the approximate formul\ae{} proposed in this section, their accuracy is very high in the practically interesting parameter range ($\lambda \leqslant 0.2$). For a given value of $\lambda$, the values of $e_{\rm dry}$ and $\tau_{\rm c}$ calculated from the closed-form expressions \eqref{eq:edry-approx} and \eqref{eq:tau-c-ansatz} were compared with the corresponding values obtained from a numerical Runge-Kutta solution of equation \eqref{eq:eom_nondim}. It was found that the expression \eqref{eq:tau-c-ansatz} for $\tau_{\rm c}$ has a maximum relative error of $2.2\times 10^{-4}$. Likewise, the relative error of the approximation \eqref{eq:edry-approx} does not exceed $4.9\times 10^{-3}$ for the values of $\lambda$ considered here. This may be seen as a numerical validation of the solutions derived in \Secref{sec:exsol:physparms}.



\section{Application to the inverse problem}
\label{sec:inversealgo}

\subsection{Inverse problem}
\noindent
In practice, one is often interested in a sort of inverse problem to the one
discussed above. The ACTM of \cite{Kempe_2012:JFM} imposes $e_{\rm dry}$ and $T_{\rm c}$ and determines the values of the parameters $k$ and $d$ such that this is obtained individually for each collision. The problem, then, is to chose the appropriate parameters \(k, d\)
depending on the incident velocity \(u_{\rm in}\), given the mass of the particle
\(m_{\rm p}\), collision time \(T_{\rm c}\) and restitution coefficient \(e_{\rm
dry}\). To the best of our knowledge, a non-numerical means of achieving this is
not known in the literature, \new{so that} \cite{Kempe_2012:JFM} \new{resorted to an iterative numerical method employing a Newton-type scheme briefly recalled below}.

\subsection{Direct determination of damping and stiffness}
\label{subsec:subroutine}
\noindent \nieuw{Equation \eqref{eq:edry-approx} allows to solve for $\lambda$, yielding
\begin{align}
\lambda & = \frac{1}{\alpha^2\tau_{\rm c,0}^2}\left(-\tfrac{1}{2}C\eta +
\sqrt{\tfrac{1}{4}C^2 \eta^2 +
\alpha^2\tau_{\rm c,0}^2\eta}\right)\:, \label{eq:lambda-reverse}
\end{align}
where $\eta = \left(\log e_{\rm dry}\right)^2$ for convenience. Only the positive root of the quadratic equation is physically relevant. The values to be used in \eqref{eq:lambda-reverse} are \(\tau_{\rm c,0} = 3.218\) \citep{Hertz1882} and \(\alpha = 1.111, C = 0.744\) from \Secref{sec:approx}.} \nieuw{When the collision time $T_{\rm c}$ is imposed in physical units, relation \eqref{eq:edry-approx} can be inserted into \eqref{eq:tau-c-ansatz} in order to calculate the unit of time}
\begin{equation}
t_{*} = \frac{T_{\rm c}}{\tau_{\rm c,0}}\sqrt{1 - A\lambda - B\lambda^2}\:,
\label{eq:t*-inv}
\end{equation}
\nieuw{where the values $A = 0.716\text{ and }B = 0.830$ from \Secref{sec:approx} are to be used.}
The \nieuw{definitions of $t_*$ and $\lambda$,} \eqref{eq:scale-lambda} and
\eqref{eq:scale-t} respectively, then yield the physical material parameters
\begin{align}
d &= \frac{2\lambda m_{\rm p}}{t_{*}}\:, \\
k &= \frac{m_{\rm p}}{\sqrt{u_{\rm in}t_{*}^5}}\:. \label{eq:ray_approximate}
\end{align}
\nieuw{For given material parameters of the particles, $m_{\rm p}$ and $e_{\rm dry}$, this provides a \new{direct, non-iterative method} to compute the parameters $k$ and $d$ in the governing equation \eqref{eq:eom} from given impact velocity $u_{\rm in}$ and desired collision time $T_{\rm c}$.}
\section{Numerical results}
\label{sec:num-test}

\subsection{Validation and assessment for binary particle-particle collisions}

\noindent In this section numerical tests of the proposed subroutine for parameter adaption \nieuw{are presented}. In particular, the accuracy \nieuw{and efficiency} of the results obtained using the \nieuw{developed} method is investigated. \Tabref{tab:accuracy-dpa} \nieuw{provides data of} test runs for various values of \(e_{\rm dry}\) \nieuw{and compares} the performance with that of the \nieuw{quasi-Newton iterations developed by \cite{Kempe_2012:JFM} to compute the desired damping and stiffness. Since the standard Newton iteration scheme requires the calculation of the Jacobi matrix for each iteration step, it was replaced by a Broyden approximation in that reference. The convergence criterion in the iterative scheme was based on the residual of the rebound velocity and collision time and was set to $10^{-6}$.} A striking feature of \nieuw{the new} algorithm is that the computation time is practically \nieuw{negligible} for all cases. This is because all the underlying calculations involve at most the evaluation of elementary mathematical functions. On the other hand, the CPU time used by the \nieuw{iterative procedure} shows a general trend of increasing CPU time for decreasing values of \(e_{\rm dry}\), and always yields \nieuw{larger} times than that required by the present \nieuw{method}.

\nieuw{Moreover,} the present subroutine does not trade off accuracy for efficiency, at least not to any significant extent. To quantify this aspect, the equations of motion were solved with the adapted parameter values \(k^{\rm appr},d^{\rm appr}\) given by the \nieuw{new} method using a third-order Runge-Kutta solver. This is similar to how the algorithm would be employed in practice, as shown in the next section. \nieuw{On this basis} the restitution coefficient \(e_{\rm dry}^{\rm appr}\) and the collision time \(T_{\rm c}^{\rm appr}\) were determined. The quality of the result can then be quantified by the relative errors 
\begin{equation}
\epsilon_{e_{\rm dry}}^{\rm appr} := \left|1 - \frac{e_{\rm dry}^{\rm appr}}{e_{\rm dry}}\right| \qquad \text{and} \qquad \epsilon_{T_{\rm c}}^{\rm appr}:= \left|1 - \frac{T_{\rm c}^{\rm appr}}{T_{\rm c}}\right|
\end{equation}
with respect to the preset target values \(e_{\rm dry} \text{ and } T_{\rm c}\). From \Tabref{tab:accuracy-dpa}, it is apparent that accuracy is indeed not compromised by the new direct method. In all cases of potential practical relevance (\(e_{\rm dry} > 0.7\)) the relative error in the restitution coefficient \nieuw{is} \(\epsilon_{e_{\rm dry}}^{\rm appr} \sim 10^{-4}\) and does not exceed $1.3 \permil$, which is excellent. For obvious reasons, the accuracy decreases when lowering \(e_{\rm dry}\). Even for \(e_{\rm dry}\) as low as \(0.4\), the relative error is \(\epsilon_{e_{\rm dry}}^{\rm appr}\sim 3 \%\), which is acceptable. In the time domain, the collision time that results from the parameter adjustment by the present method agrees up to a relative error \(\epsilon_{T_{\rm c}}^{\rm appr} \sim 10^{-4}\) with the pre-set target value.

\nieuw{The computation times for the Newton iteration scheme given in \Tabref{tab:accuracy-dpa} were obtained for a much more stringent convergence criterion ($10^{-6}$, as stated above). Hence, one may ask what the computational effort is, if the required accuracy of the Newton scheme is reduced to that of the direct approximate scheme, simply by stopping the iterations earlier. The resulting timings are reported in \Tabref{tab:newtimings}. It turns out that in the extreme case of $e_{\rm dry} = 0.4$, the CPU time required when relaxing accuracy is reduced by a factor of nearly 6.3\,. For $e_{\rm dry} = 0.95$ used later in \Secref{sec:numtest:gross}, the \new{CPU time is practically not reduced at all} when lowering the accuracy of the iterative scheme. In all cases, even with the relaxed convergence criterion, the required time is orders of magnitude higher than the direct scheme, because even the initialisation and execution of 4-5 iteration steps is much costlier than the evaluation of elementary functions.}

In summary, it may be concluded that the method proposed in this paper enables a computationally cheap and \nieuw{at the same time} sufficiently accurate implementation of the concept of parameter-adapted time-stretched collision modelling \nieuw{developed by} \cite{Kempe_2012:JFM}.

\subsection{Application to multiple particles sedimenting on a rough surface}
\label{sec:numtest:gross}
\noindent
The final test case reported here deals with the sedimentation of 100 randomly distributed particles and their impact on a layer of 195 fixed particles arranged in hexagonal packing (Fig.~\ref{fig:100_on_hex_sketch}a) very similar to the simulations presented by \cite{kempe:2014_IJMPF}. The main difference is that the fluid is neglected in the present case corresponding to an infinite Stokes number of the particles. 
The goal of the simulations is to compare the results obtained with the original scheme of \cite{Kempe_2012:JFM} \new{when applied to this case with infinite Stokes number,} to the results of the approximate inverse procedure developed in \Secref{sec:inversealgo}. \new{The situation is very close to practical applications as it features multiple simultaneous collisions.} \nabo{At this point, it should be recalled that, in keeping with the pre-declared scope
of the present work, the simulation presented below only considers normal collision forces,
since only these enter the computations of the ACTM.
In other words, the collisions are generally oblique, but with
the tangential forces supposed to be negligibly small.} The goal of the test is to demonstrate sufficiently close agreement of the results obtained using the direct algorithm with those of the original iterative procedure and futhermore to obtain timings for such a realistic situation.

The computational domain is $\Omega = [0, L] \times [0, L ] \times [0, L]$ with $L = 1.5$. Periodic boundary conditions were applied in the $x$- and $z$-direction. The  gravitational acceleration is $g = 9.81$ and the density of the particles is $\varrho_{\rm p} = 1200$, with the particle diameter $D = 0.1154$. The mobile particles are placed randomly in the subdomain $\Omega_1 = [0.1, 1.4] \times [0.3, 1.2] \times [0.1, 1.4]$\nieuw{, and their} velocity initialized with zero. 
Once \nieuw{these} particles are released from their initial position (\Figref{fig:100_on_hex_sketch}a), they are accelerated by gravity towards the fixed layer and then collide with the bed or with other mobile particles. At the same time they are subjected to a dissipation of kinetic energy \nieuw{when undergoing collisions}. Two different cases are considered here. In Case~1, the coefficient of restitution is $e_{\rm dry}=0.95$ and in Case~2 the \nieuw{value} is $\nieuw{e_{\rm dry}} = 0.7$. The simulations were run with $\Delta t = 5\times10^{-4}$ for 5000 steps\nieuw{,} corresponding to a non-dimensional simulation time of $t=2.5$\,. This situation is depicted in \Figref{fig:100_on_hex_sketch}b for Case 1. The particles are coloured according to the absolute value of their velocity $u_{\rm p} = |{\bf u}_{\rm p}|$ from red ($u_{\rm p}=2$) to blue ($u_{\rm p}=0$) showing that not all the particles come to rest at the end of the simulation if the damping is weak.

The collision process is elucidated by computing the various components of the total energy of the particles. The potential energy, the kinetic energy and the energy stored by deformation of the springs in the collision model are defined as
\begin{equation}\label{eq:e_pot}
	E_{\rm pot}= \sum_{i=1}^{n_p} m_{p,i} \:  g  \: y_{p,i}
\end{equation}
\begin{equation}\label{eq:e_kin}
E_{\rm kin}= \frac{1}{2} \: \sum_{i=1}^{n_p}  m_{p} \left| \textbf{u}_{p,i}\right|^2 
\end{equation}
and
\begin{equation}\label{eq:e_spring}
E_{\rm spring}= \sum_{i=1}^{n_p} \int F_c \left|\zeta_{n,pq}\right| \: \text{d} \zeta_{n,pq} = \frac{5}{2} \:  \sum_{i=1}^{n_p} k_n \, \left|\zeta_{n,pq}\right|^{\,5/2} \hspace{0.25cm},
\end{equation}
respectively.
The total energy in the computational domain then is
\begin{equation}\label{eq:e_total}
	E_{\rm tot}= E_{\rm pot} + E_{\rm kin} + E_{\rm spring} \hspace{0.25cm}.
\end{equation}

The fractions of energy are displayed in \Figref{fig:100_on_hex_sketch_energy}a and \Figref{fig:100_on_hex_sketch_energy}b for Case 1 and 2, respectively. As already mentioned above, in Case 1 the particles \nieuw{do} not come to rest at the end of the simulation, reflected by their kinetic energy not being zero at the end of the simulation. In contrast to this, the kinetic energy of the particles \nieuw{at $t = 2.5$} is zero for Case 2.
Obviously, no significant differences of the various fractions of the energy are observed if the iterative or if the approximate method is used. \nieuw{The slight differences that are seen in Case 1 rather result from the fact that extremely small differences in the collision process can yield different particle trajectories, in turn leading to subsequent differences, as in a billiard system, without changing the overall bulk behaviour of the ensemble.} 

To further elucidate the efficiency of the two numerical procedures, the CPU times of the \nieuw{old} and the \nieuw{new} scheme are \nieuw{compiled} in \Tabref{tab:times_actm}.
Here, $t_{\rm tot}$ is the overall CPU time required for the whole simulation \nieuw{including overhead}, $t_{\rm par}$ the time spen\nieuw{t} in the particle routines, $t_{\rm coeff}$ is the part of $t_{\rm par}$ required for the determination of stiffness and damping, $n_{\rm col}$ the overall number of collisions and, finally, $t_{\rm col}$ is the average time required per collision. Obviously, the numerical effort is substantially reduced for all values of $e_{\rm dry}$ if the new approximate method is used. The results presented in this section confirm the accuracy, robustness and efficiency of the proposed method.

\section{Concluding remarks}
\label{sec:discus}
\noindent
In this paper, normal particle-wall and particle-particle collisions were studied \nieuw{by modelling} the elastic interaction \nieuw{with} a repulsive Hertzian contact force and an additional damping force linear in the velocity.

First, the equation of motion \eqref{eq:eom} was converted to its dimensionless form \eqref{eq:eom_nondim}. This reduces the number of parameters in the equation to a single constant \(\lambda\) depending on the material parameters \(k,d\) and the impact velocity \(u_{\rm in}\). In particular, the physically relevant and practically interesting properties, the collision time \(T_{\rm c}\) (dimensionless: \(\tau_{\rm c}\)) and restitution coefficient \(e_{\rm dry}\), hence, only depend on this particular combination of physical input parameters. This is conceptually and technically analogous to the damping ratio, or its inverse, the quality factor, which are familiar from the simple harmonic oscillator, the latter more so from $LC$ circuits in electrical engineering but appears to be new in the present setting of damped Hertzian collisions. In contrast, earlier works \citep{KruggelEmden2007,Kempe_2012:JFM} typically use a three-parameter family of input variables $k,d,u_{\rm in}$. While the use of $e_{\rm dry}$ to label cases is helpful in practice, the proposed nondimensionalisation and the subsequent analytical investigation establishes in a straightforward manner the one-to-one correspondence between the important experimental parameter $e_{\rm dry}$ and the parameter $\lambda$, which is easy to obtain from material parameters. Furthermore, it demonstrates that different parameters with the same $e_{\rm dry}$ have solutions that are identical up to a linear scaling of space and time.

Next, an exact formal solution of the governing equation using nonlinear transformations and a series in the parameter \(\lambda\) was proposed, \nieuw{providing} a rigorous calculation of \(e_{\rm dry}\) and \(\tau_{\rm c}\) from the equation of motion. Owing to the technical difficulties presented by the governing equation, the analysis is necessarily involved, the methods and technique employed may be instructive, \nieuw{and} the results may be of use in theoretical considerations. The approach is also quite flexible, and a similar analysis would be applicable to more general settings,
%
%
such as the extension to fully nonlinear models, i.e. \nieuw{a} nonlinear spring force in combination with a nonlinear dissipative force \nieuw{as considered in} some studies \citep{kawabara1987restitution,crowe:2006,crowe:2002}. This may be the subject of future investigation.

Subsequently, \nieuw{an approximation based on the exact solution was proposed, yielding} compact and convenient formul\ae{} for \(e_{\rm dry}\) and \(\tau_{\rm c}\). Inverse formul\ae{} were derived on that basis, enabling a very efficient and accurate calculation of the required stiffness and damping from the given input parameters, i.e. the desired collision time $T_{\rm c}$ and restitution coefficient $e_{\rm dry}$. These were then applied to \nieuw{an} engineering context.
The \nieuw{quasi-Newton iteration} scheme in the \nieuw{original} ACTM \nieuw{was} replaced by the direct approximate solution developed here. Numerical tests on binary and multiple particle collisions confirm the accuracy and efficiency of the proposed method.

\section*{Acknowledgements}
\noindent The computations in \Secref{sec:num-test} were performed at ZIH, TU Dresden. The authors thank Prof. Dr. Ralph Chill and Prof. Dr. J\"urgen Voigt for helpful discussions on an early status of this work. The first author wishes to express his gratitude to the Martin-Andersen-Nex\"o-Gymnasium Dresden for providing the appropriate framework within its school curriculum to carry out parts of the research.

\clearpage\newpage
\appendix
\makeatletter
\gdef\thetable{\@arabic\c@table}
\makeatother
\makeatletter
\gdef\thefigure{\@arabic\c@figure}
\makeatother
\section{Physical interpretation of the coefficients \(v_{\pm,m}\)}
\label{app:PhysInt}
\noindent
\nieuw{I}t may be instructive to \nieuw{discuss} the physical meaning of the coefficients \(v_{\pm,m}\) in the series expansion in powers of \(\lambda\) of \(v_{\pm}\), equations \eqref{eq:v+m-master}--\eqref{eq:v-master}. First, it is readily seen that \nieuw{the zeroth-order term is}
\[
v_{+,0}=v_{-,0}=1 - \tfrac{4}{5}y^5\;,
\]
as obtained in \eqref{eq:zerothorder}. The fact that the two expressions are equal indicates reversibility. Indeed, for \(\lambda=0\), the classical oscillation problem with Hertzian restoring force is obtained, which is a conservative system. Moreover, upon inspection of the functional dependence, it is evident that the zeroth-order term expresses conservation of mechanical energy itself. \nieuw{T}he kinetic energy per unit mass and in dimensionless form is given by 
\[
E_{\rm kin} = \tfrac{1}{2}\dot{z}^2 = \tfrac{1}{2}v_{\pm}\;,
\]
depending on which part of the trajectory is considered. Thus, the auxiliary variables \(v_{\pm}\) are proportional to the kinetic energy. The potential energy associated with the Hertz contact force (dimensionless and per unit mass) is 
\[
E_{\rm pot} = \tfrac{2}{5}z^{5/2} = \tfrac{2}{5}y^5\;.
\]
For both the inward and outward motion of the particle, the conservation of energy
\begin{equation}
\tfrac{1}{2}v_{\pm} + \tfrac{2}{5}y^5 = E_{\rm kin} + E_{\rm pot} = {\rm const.}
\end{equation}
is valid to zeroth-order in $\lambda$. 

\nieuw{N}ow consider the more involved first-order corrections. \nieuw{Equations} \eqref{eq:v+m-master}--\eqref{eq:v-master} \nieuw{yield}
\begin{align}
v_{+,1} &= -8\int_{0}^{y}\!\sqrt{1 - \tfrac{4}{5}y^5}\,y^\prime{\rm d}y^\prime \\
v_{-,1} &= \left.\frac{\partial e_{\rm dry}^2}{\partial \lambda}\right|_{\lambda=0} + 8\int_{0}^{y}\! \sqrt{1 - \tfrac{4}{5}y^5} \,{\rm d}y^\prime y^\prime\;.
\label{eq:vpm1}
\end{align}
Both integrals can be evaluated analytically to give 
\begin{equation*}
v_{+,1} = -\tfrac{16}{9}y^2\sqrt{1 - \tfrac{4}{5}y^5} - \tfrac{20}{9}y^2\,{}_2F_1\!\left(\tfrac{2}{5},\tfrac{1}{2};\tfrac{7}{5};\tfrac{4}{5}y^5\right)
\end{equation*}
with \({}_2F_1\) denoting the hypergeometric function \citep[cf.][Chapter 15, pp. 555]{AbraSteg}. A similar expression results for $v_{-,1}$. The integral representations are actually more conducive for physical interpretation.
Indeed, both expressions can be interpreted as work done by the dissipative force. In case of \(v_{-,1}\), there is an added constant term, because the initial condition \(v_{-}(y=0)=e_{\rm dry}^2\) depends on \(\lambda\), unlike the much more straightforward case \(v_{+}(y=0)=1\). For this reason, the +, or inward, branch of the trajectory \nieuw{shall be discussed} in the following. Since \(2ydy=dz\), 
\begin{align}
\tfrac{1}{2}v_{+,1} &= -2\int_{0}^{z}\! \sqrt{1 - \tfrac{4}{5}z^{5/2}}\,{\rm d}z^\prime\;.
\end{align}
On the other hand, from \(v_{+} = v_{+,0} + \mathcal{O}(\lambda)\), it follows that to first order in \(\lambda\), the work done by the damping force is
\begin{equation}
W_{+}^{\rm diss} = -2\lambda\int_{0}^{z}\! \dot{z}\,{\rm d}z^\prime = -2\lambda\int_{0}^{z}\! \sqrt{v_{\pm,0}}\,{\rm d}z^\prime + \mathcal{O}(\lambda^2) = \tfrac{1}{2}v_{+,1}\lambda + \mathcal{O}(\lambda^2)\;.
\end{equation}
Thus, \(v_{\pm,1}\) are essentially first-order corrections in the energy balance due to the work done by the dissipative term in the equations of motion.

\clearpage

\setlength{\unitlength}{1cm}
\begin{figure}
\begin{picture}(7,7.7)
    \put(0.5,0.2){\includegraphics[width=0.57\textwidth]{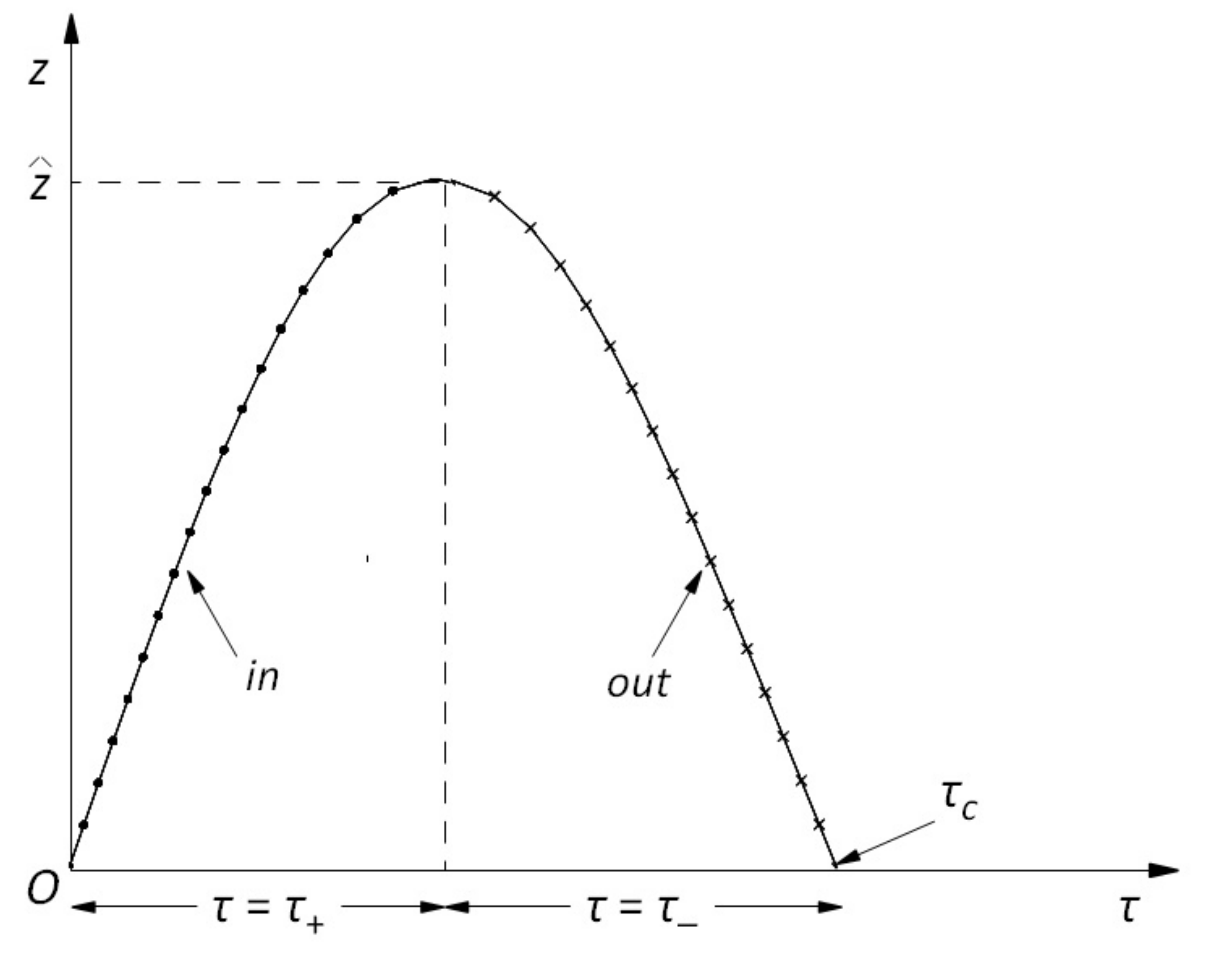}}
%
    \put(10.0,0.2){\includegraphics[width=0.4\textwidth]{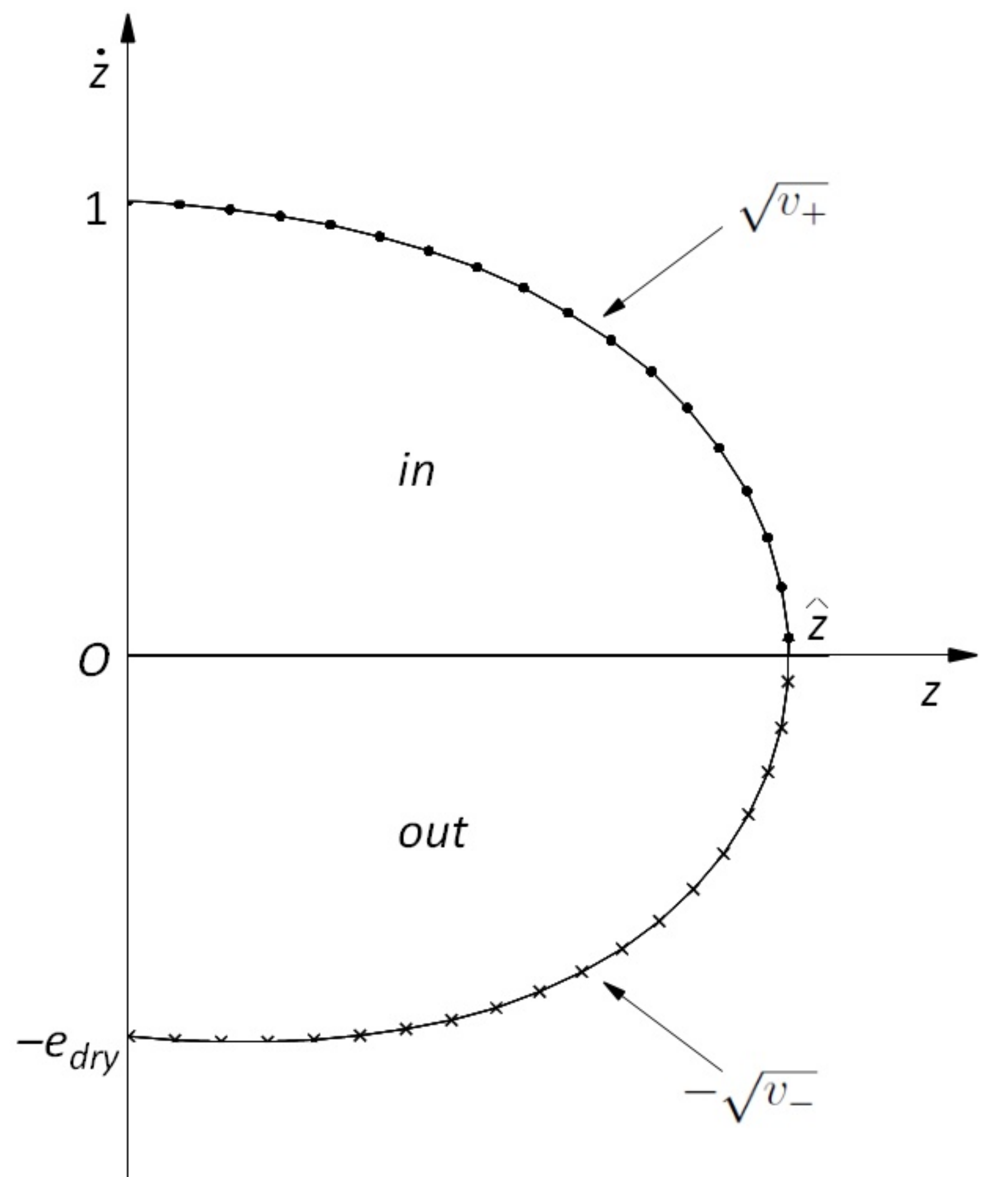}}   
%
    \put(5.0,-0.0)     {\emph{a}) }
    \put(13,-0.0)     {\emph{b}) }
\end{picture}
    \caption{{Sketch of the compression (in) and the rebound phase (out) during the collision process (cf. Figure \ref{fig:sketch_collision}) with $\hat{z}$ being the non-dimensional maximum surface penetration. a) Physical space, b) Phase space }}
    \label{fig:zt_ph} 
\end{figure}

\setlength{\unitlength}{1cm}
\begin{figure}[ht]
\begin{picture}(7,7)
    \put(0,0){\includegraphics[width=0.45\textwidth]  {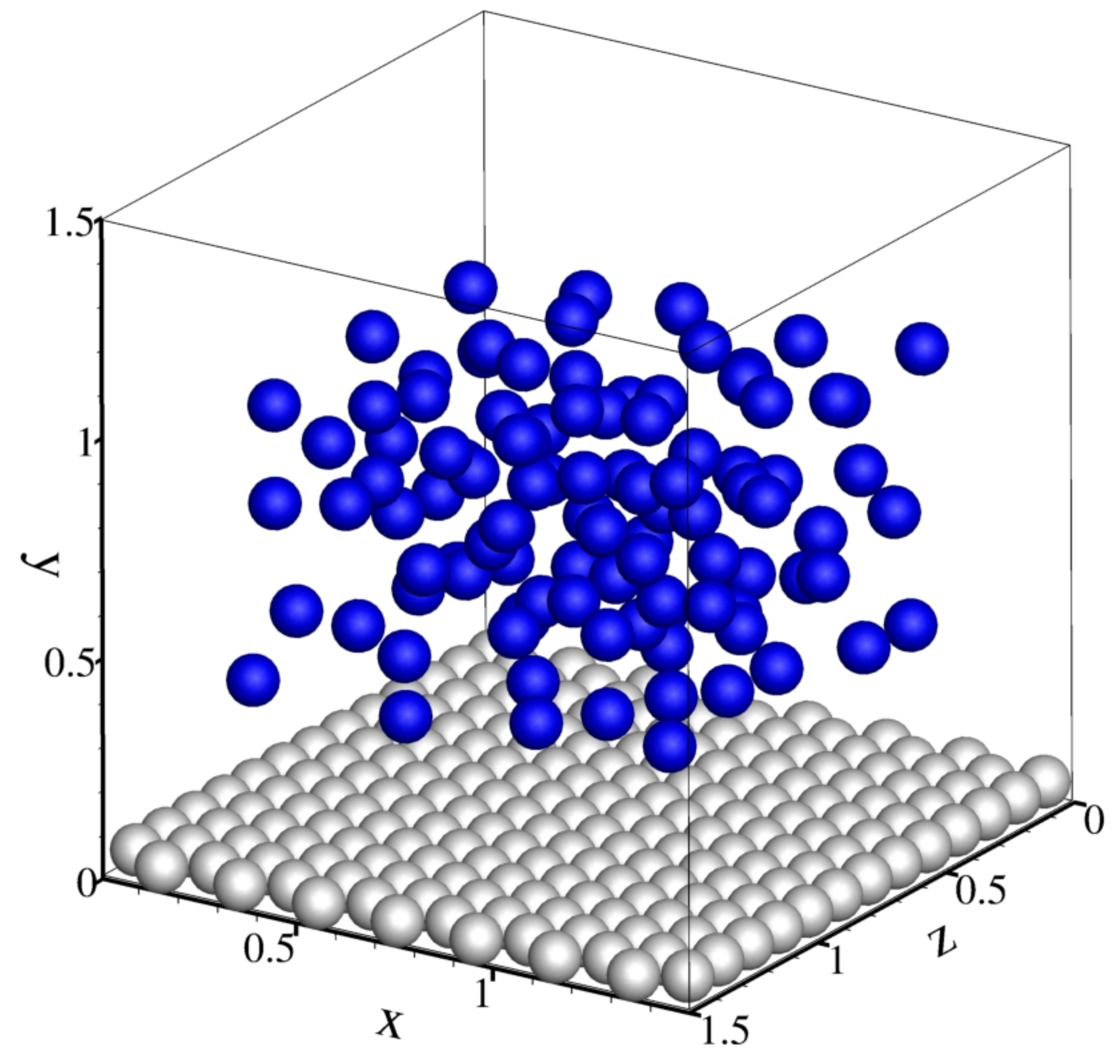}}
    \put(8.5,0){\includegraphics[width=0.45\textwidth]{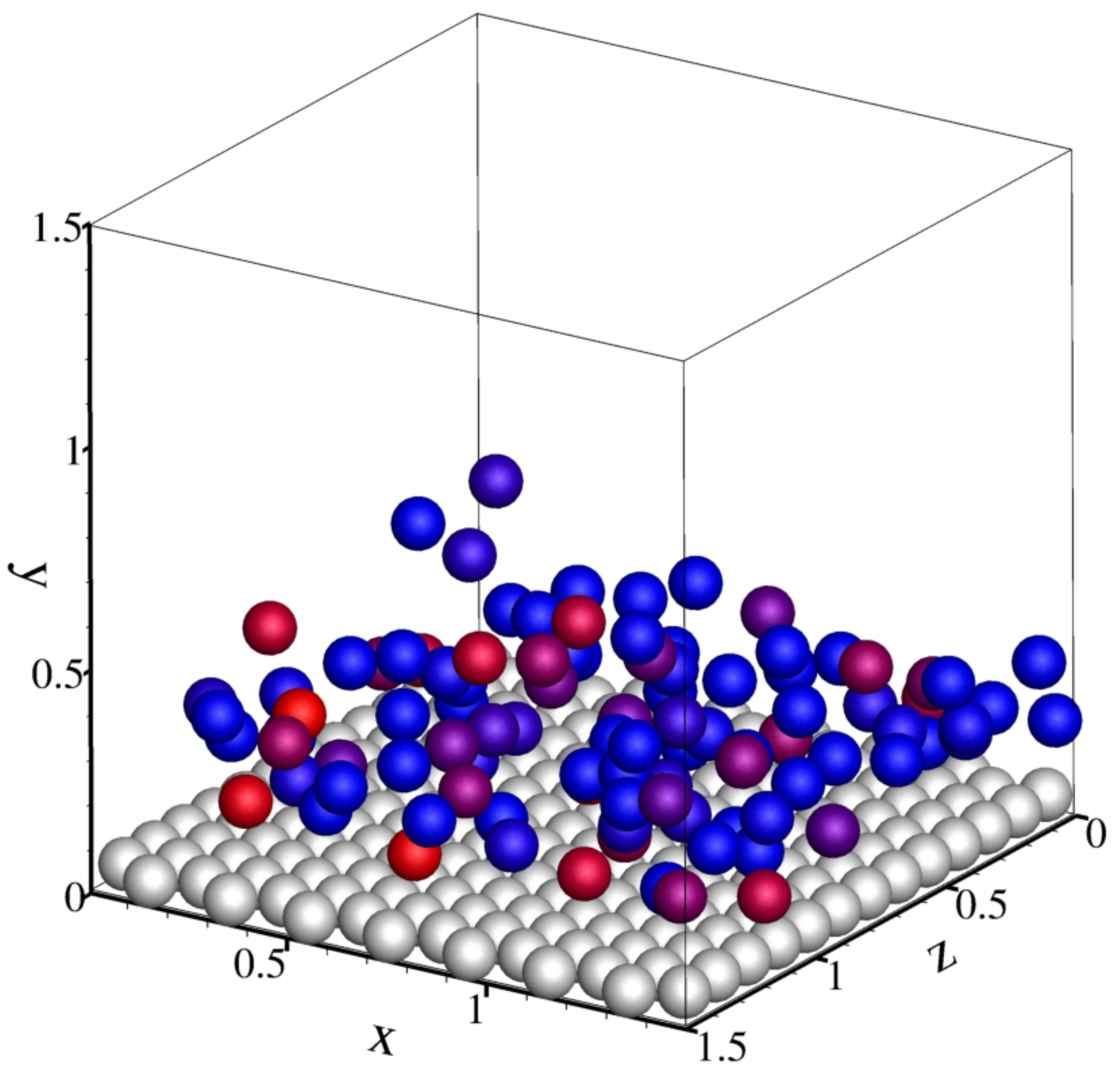}}
  \put(0.5,0.2)     {a) }
  \put(8.9,0.2)     {b) }
\end{picture}
     \caption{{Sedimentation of 100 particles at infinite Stokes number and impact on a layer of 195 fixed particles in hexagonal packing with $e_{\rm dry} = 0.95$ (Case 1). a) Initial configuration at $t=0$, b) situation at the end of the simulation $t=2.5$. The particles are coloured by the absolute value of their velocity from red ($u_p=2$) to blue ($u_p=0$).}}
     \label{fig:100_on_hex_sketch} 
\end{figure}
\setlength{\unitlength}{1cm}
\begin{figure}[ht]
\begin{picture}(7,7)
    \put(1,0){\includegraphics[width=0.4\textwidth]   {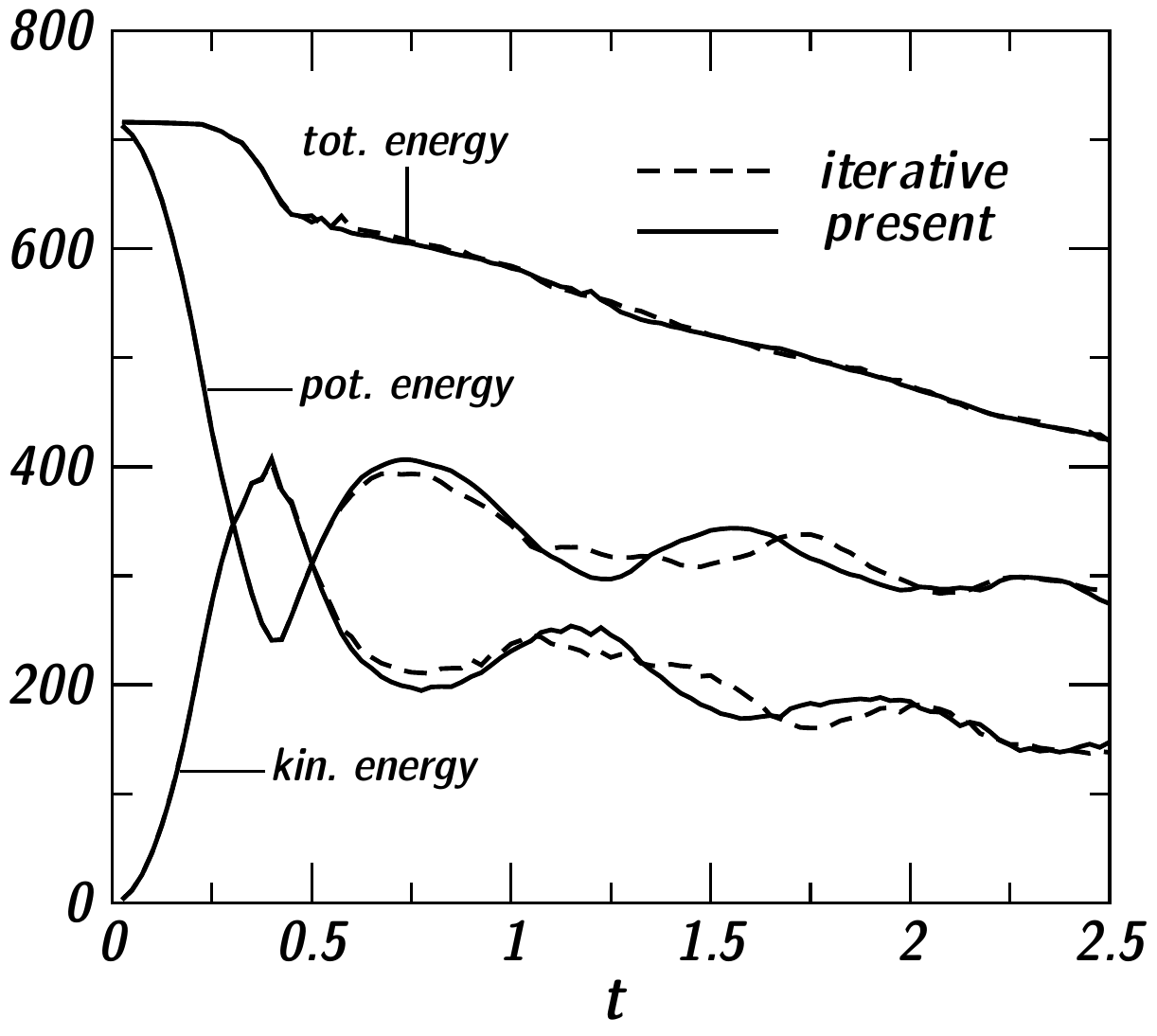}}
    \put(9,0)   {\includegraphics[width=0.4\textwidth]{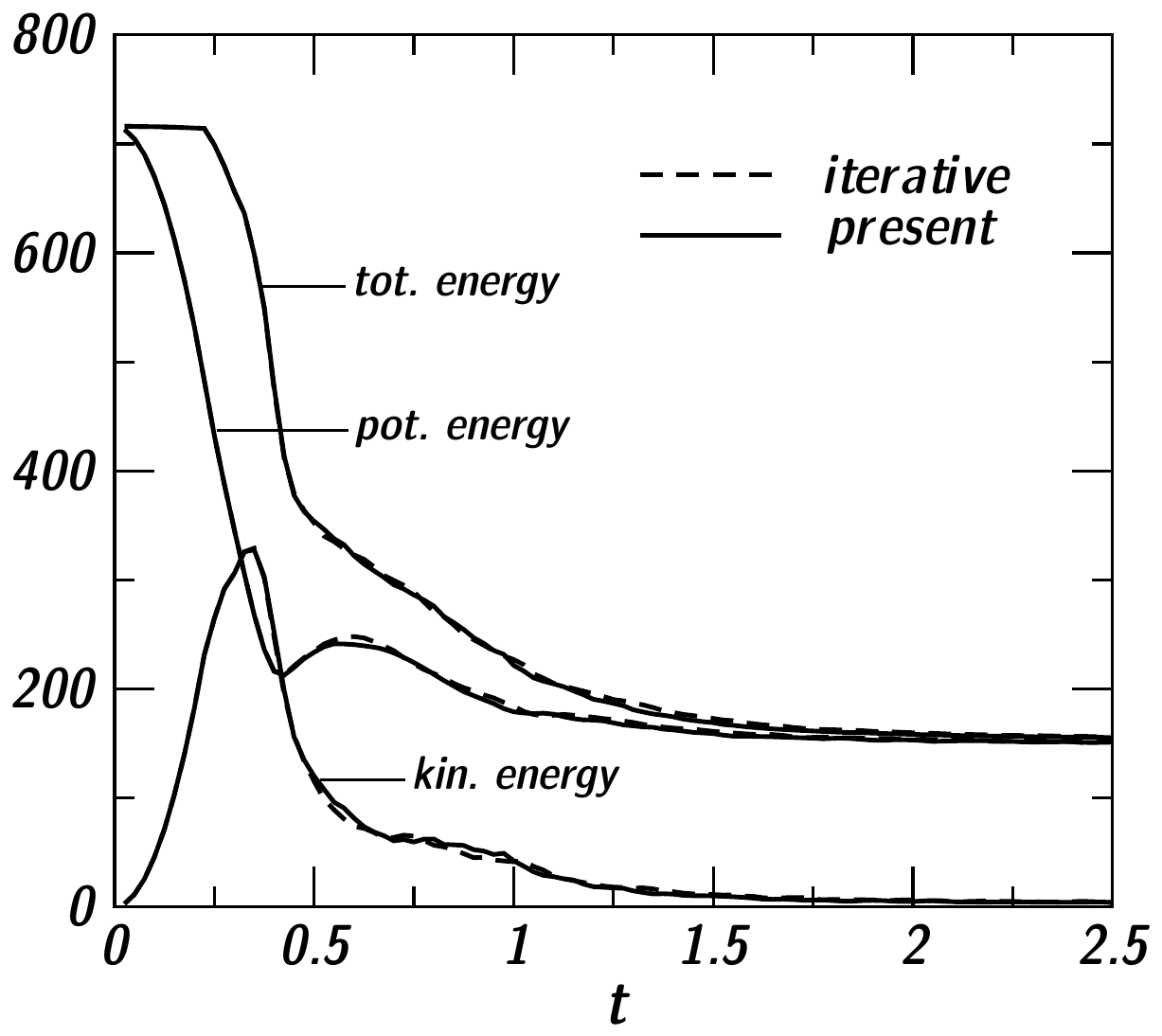}}
  \put(0.5,  4.0)   {a) }
  \put(8.5,4.0)     {b) }
\end{picture}
     \caption{{Fractions of energy for the sedimentation of 100 mobile particle and collision with a layer of fixed particles using the ACTM \citep{Kempe_2012:JFM}, labelled `iterative', and the present approximate scheme \eqref{eq:lambda-reverse}--\eqref{eq:ray_approximate}. (a) Case 1 with $e_{\rm dry}=0.95$, (b) Case 2 with $e_{\rm dry}=0.70$.}}
     \label{fig:100_on_hex_sketch_energy}
\end{figure}
\clearpage

\newgeometry{top=0pt, bottom=0pt}
\begin{landscape}
\begin{table}
\centering
\begin{tabular}{|r|rcllll|lllllll|}
\hline  
\multirow{2}*{\(e_{\rm dry}\)} & \multicolumn{6}{|c|}{Newton iterations} & \multicolumn{7}{|c|}{present} \bigstrut \\
\cline{2-14}
 & \(n_{\rm it}\) & \(t_{\rm CPU}\) [s] & \(k\) & \(d\) & \(u_{\rm out}\) & \(T_{\rm c}\) & $t_{\rm CPU}$ [s] & \(k\) & \(d\) & \(u_{\rm out}\) & \(T_{\rm c}\) & \(\epsilon_{e_{\rm dry}}\) & \(\epsilon_{T_{\rm c}}\) \bigstrut \\
\hline
\adaptparm{1.00}{1}{---}{60699.0}{0}{\(-1\)}{\(10^{-2}\)} & $2.76 \times 10^{-7}$ \result{\(60694.4\)}{0.00000}{\(-1.00000\)}{\(1.00000 \times 10^{-2}\)}{$3.50 \times 10^{-12}$}{0} \bigstrut[t]\\
\adaptparm{0.95}{5}{0.0234}{61491.7}{0.30165}{\(-0.95\)}{\(10^{-2}\)} & $2.88 \times 10^{-7}$ \result{61491.7}{0.30165}{\(-0.95000\)}{\(1.00001\times 10^{-2}\)}{\(7.71\times 10^{-7}\)}{\(7.6\times 10^{-5}\)}\\
\adaptparm{0.90}{5}{0.0234}{62369.2}{0.61964}{\(-0.90\)}{\(10^{-2}\)} & $2.86 \times 10^{-7}$ \result{62371.9}{0,61966}{\(-0.90000\)}{\(0.99998\times 10^{-2}\)}{\(1.33\times 10^{-5}\)}{\(1.7\times 10^{-5}\)}\\
\adaptparm{0.80}{13}{0.0507}{64418.6}{1.31209}{\(-0.80\)}{\(10^{-2}\)} & $2.86 \times 10^{-7}$ \result{64437.6}{1.31340}{\(-0.79979\)}{\(0.99991\times 10^{-2}\)}{\(2.63\times 10^{-4}\)}{\(9.5\times 10^{-5}\)}\\
\adaptparm{0.70}{8}{0.0351}{66985.7}{2.09541}{\(-0.70\)}{\(10^{-2}\)} & $2.87 \times 10^{-7}$ \result{67042.7}{2.10348}{\(-0.69906\)}{\(0.99982\times 10^{-2}\)}{\(1.34\times 10^{-3}\)}{\(1.8\times 10^{-4}\)}\\
\adaptparm{0.60}{19}{0.0646}{70299.5}{2.99798}{\(-0.60\)}{\(10^{-2}\)} & $2.82 \times 10^{-7}$ \result{70434.0}{3.02380}{\(-0.59734\)}{\(0.99979\times 10^{-2}\)}{\(6.05\times 10^{-3}\)}{\(2.1\times 10^{-4}\)}\\
\adaptparm{0.50}{60}{0.1679}{74735.8}{4.06030}{\(-0.4999\)}{\(10^{-2}\)} & $2.75 \times 10^{-7}$ \result{75047.0}{4.12956}{\(-0.49396\)}{\(1.00000\times 10^{-2}\)}{\(1.21\times 10^{-2}\)}{\(2.5\times 10^{-6}\)}\\
\adaptparm{0.40}{76}{0.1992}{80999.5}{5.35192}{\(-0.3998\)}{\(10^{-2}\)} & $2.88 \times 10^{-7}$ \result{81738.6}{5.51951}{\(-0.38797\)}{\(1.00085\times 10^{-2}\)}{\(1.34\times 10^{-2}\)}{\(8.5\times 10^{-4}\)} \bigstrut[b]\\
\hline
\end{tabular}
\captionsetup{width=1.75\textheight}
\caption{Summary of test runs \nieuw{for} binary particle-particle collisions. The results achieved by the method presented in \Secref{subsec:subroutine} (heading: present) are compared with those employing the \nieuw{Newton iterations} proposed by \cite{Kempe_2012:JFM}. The situation simulated is depicted in \Figref{fig:sketch_collision} with initial condition \(u_{\rm in}=1\), particle density \(\varrho=7800\) and radius \(R_{\rm p} = 10^{-2}\). The pre-set collision time was \(T_{\rm c} = 10^{-2}\) and the target restitution coefficient \(e_{\rm dry}\) was varied as documented in the first column. The cases with \(e_{\rm dry} < 0.7\) are fictitious, in the sense that they are unlikely to be of practical interest as of now, and are included to show the integrity of the presented method even in such extremes.}
\label{tab:accuracy-dpa}
\end{table}
\end{landscape}
\restoregeometry

\begin{table}
\centering
\begin{tabular}{|l|cc|}
\hline
$e_{\rm dry}$ & $n_{\rm it}^\prime$ & $t_{\rm CPU}^\prime$ [s] \bigstrut \\
\hline
1    & --- & ---    \bigstrut[t] \\
0.95 & 4   & 0.0234 \\
0.90 & 4   & 0.0234 \\
0.80 & 4   & 0.0234 \\
0.70 & 4   & 0.0234 \\
0.60 & 4   & 0.0234 \\
0.50 & 5   & 0.0313 \\
0.40 & 5   & 0.0313 \bigstrut[b] \\
\hline
\end{tabular}
\caption{Number of iterations $n_{\rm it}^\prime$ and computation time $t_{\rm CPU}^\prime$ for the quasi-Newton iteration scheme of \cite{Kempe_2012:JFM} with the convergence criterion at par with the accuracy achieved by the present method (cf. \Tabref{tab:accuracy-dpa}).}\label{tab:newtimings}
\end{table}

\begin{table}
\centering
\begin{tabular}{| l | l c c c c c |}
\hline 
Method & $e_{\rm dry}$ &  $ t_{\rm tot} $ [s] &  $ t_{\rm par}$ [s]  &  $t_{\rm coeff}$ [s] & $n_{\rm col}$ & $t_{\rm col}$ [s] \bigstrut \\
\hline
Newton & 0.95     & 836.46       &   812.78     &    221.49    &  21950    & 0.01009      \bigstrut[t] \\
 & 0.90     & 839.22       &   815.23     &    229.71    &  22362    & 0.01027      \\
 & 0.80     & 783.91       &   760.36     &    186.57    &  16605    & 0.01134      \\
 & 0.70     & 734.28       &   711.93     &    142.86    &  11687    & 0.01222      \bigstrut[b] \\
 \hline
present & 0.95     & 597.76     &   578.45     & $1.953 \times 10^{-2}$  &  21111    &  $9.72 \times 10^{-7}$    \bigstrut[t] \\
 & 0.90     & 601.75     &   582.09     & $3.516 \times 10^{-2}$  &  22974    &  $9.05 \times 10^{-7}$    \\
 & 0.80     & 591.66     &   572.28     & $2.344 \times 10^{-2}$  &  18845    &  $8.32 \times 10^{-7}$    \\
 & 0.70     & 591.19     &   571.71     & $7.813 \times 10^{-3}$  &  11829    &  $8.38 \times 10^{-7}$    \bigstrut[b] \\
\hline 
\end{tabular}
 \caption{{CPU time required for the sedimentation test case with various values of the restitution coefficient using the ACTM \nieuw{with Newton iterations} \citep{Kempe_2012:JFM} and the present direct method \eqref{eq:lambda-reverse}--\eqref{eq:ray_approximate} for the computation of stiffness and damping. Nomenclature defined in the text.}}
 \label{tab:times_actm}
\end{table}
\end{document}